\documentclass[runningheads]{llncs}
\usepackage[T1]{fontenc}
\usepackage{newtxtext}       %
\usepackage[varvw]{newtxmath}       

\newcommand{\pro}{\textcolor{green!55!black}{$\oplus$}~}
\newcommand{\con}{\textcolor{red!70!black}{$\ominus$}~}

\usepackage{graphicx}

\makeatletter
\newcommand{\pattern}[2]{%
\section*{#2}
\label{#1}
}

\makeatother
\newcommand{\providepatname}[2]{\expandafter\gdef\csname patname:#1\endcsname{#2}}
\makeatletter
\newcommand{\patref}[1]{%
  \hyperref[pat:#1]{\@ifundefined{patname:#1}{#1}{\csname patname:#1\endcsname}}%
}
\makeatother
\newcommand{\field}[1]{\par\smallskip\noindent\textbf{#1.}\hspace{0.4em}\ignorespaces}

\begin{document}
\title{Patterns for Designing CI/CD Workflows}
%
%
\author{Leandro Sousa\inst{1} \and
Filipe Correia\inst{1}}
\authorrunning{L. Sousa and F. Correia}
%
\institute{INESC TEC, Faculty of Engineering, University of Porto, Portugal\\
\email{\{up202300371,filipe.correia\}@fe.up.pt}}
\maketitle              

\pattern{fail-fast}{Fail Fast}

\field{Context}  A pipeline composed of multiple interdependent phases/steps.

\field{Problem} In the event of a failure at any stage of a pipeline, the outcome of the workflow is predetermined. However, the pipeline continues to execute subsequent steps, the results of which are contingent on the failed step. In the event of a failure, how should the pipeline behave?

\field{Forces}
\begin{itemize}\setlength\itemsep{1pt}\setlength\topsep{2pt}
  \item \emph{Early Feedback vs Failure Visibility} --- Aborting the pipeline after a failure reduces time and resources, but may hide other failures that maybe revealed later in the execution.
  \item \emph{Diagnostic Completeness vs Results Validity --- The goal is to derive maximum diagnostic value from each workflow execution, yet the steps that depend on a failed step may run on invalid or non-existent inputs, so the result information might be misleading, contributing to noise rather than information.}
  \item \emph{Diagnostic Completeness vs Results Validity} --- A pipeline execution should run all steps that have the potential to reveal a problem; however, it should be noted that execution time and executor capacity are both finite and shared across the team, or organization. Thus, the execution of a workflow that has already been identified as failed will delay the progress of all other tasks.
\end{itemize}

\field{Solution} Design your pipeline so that, when a part of it fails, other dependent parts are not executed and automatically fail instead~\cite{bouzenia2024resource,duvall2007continuous}.

\field{Consequences}
\begin{itemize}\setlength\itemsep{1pt}\setlength\topsep{2pt}
  \item[\pro] Developers can catch the failures earlier has they receive feedback sooner.
  \item[\pro] There is no wasted time and resource in the dependent steps.
  \item[\con] Latent defects might remain undetected until further pipeline executions.
\end{itemize}

\field{Related patterns} The \emph{Single-Responsibility Steps} is a key Pattern that enables this Pattern effectiveness as stopping for failing fast a workflow can be more precise that the granularity of the steps themselves. The use of specific, smaller steps help to reveal the independence between checks which enables this pattern. Combining the two comes at a cost: finer steps mean a larger dependency graph must be declared and maintained. Applying first \emph{Single-Responsibility Steps} enable a better implementation of \emph{Fail Fast}.

\pattern{split-workflow}{Single-Responsibility Steps}

\field{Context}  A pipeline workflow uses custom workflow definition maintained and authored by a specific team.

\field{Problem} A step may be the most granular part of a workflow. Thus, encompassing unique name, failure point, retry mechanism, caching and reporting mechanisms. A step that performs multiple actions report a single result while a workflow with split steps solve this it carries its own declaration and setup cost. How can we determine the work a step should do?~\cite{gallaba2020use}

\field{Forces}
\begin{itemize}\setlength\itemsep{1pt}\setlength\topsep{2pt}
  \item \emph{Diagnostic simplicity vs Orchestration complexity} --- It is imperative that failures are immediately apparent and straightforward to find, thus encouraging the division of the workflow into single-responsibility steps; However the addition of further jobs and steps with their own declaration can make this decomposition that make failures easily found make the workflow progressively hard to read and change.
  \item \emph{Step isolation vs State continuity} --- It is imperative that the steps are executed in an independent manner, thus ensuring that a potentially unreliable step does not compromise the integrity of the whole workflow. However, independent steps remove the implicit shared state of a single script. Thus, every data that needs to be shared across steps must be passed explicitly.
\end{itemize}

\field{Solution} Try to make each step in a workflow as simple and self evident as possible by limiting the number of commands it executes to the essentials, as well as naming the step if necessary~\cite{khatami2024catching,duvall2007continuous,gallaba2020use}.

\field{Consequences}
\begin{itemize}\setlength\itemsep{1pt}\setlength\topsep{2pt}
  \item[\pro] Clear and simple steps makes easier to understand the CI/CD workflow.
  \item[\pro] By breaking pipeline tasks down into specific steps, a pipeline failure is transformed into easy locatable steps.
  \item[\con] Breaking a pipeline into small specific steps can result into spread configurations which can decrease the overall pipeline execution time.
\end{itemize}

\field{Related patterns} \emph{Fail Fast} make this Pattern worthwhile. Creating single-responsibility steps improves how can maintainers control how the pipeline behave, combining this with \emph{Fail Fast} skips the downstream steps after a failure.

\pattern{iopr}{Interrupt Outdated Pipeline Runs}

\field{Context}  Pipelines where only their most up to date results matter.

\field{Problem} Teams and developers pushes multiple times in a short period of time causing multiple pipeline executions. The first run is still execution when the last arrives so there are concurrently executions. The code introduced before the last execution has already been replaced, thus it is a wast of time and computational resources to continue a pipeline run when its input is outdated. Outdated pipeline results are irrelevant. A pipeline's result is outdated when the input used to generate that result is older than the most recent input~\cite{bouzenia2024resource}.

\field{Forces}
\begin{itemize}\setlength\itemsep{1pt}\setlength\topsep{2pt}
  \item \emph{Increase available capacity vs Discard completed work} --- If a run is superseded we want to reclaim the runner capacity as fast as possible. Nevertheless, interruption a pipeline in a mid-way step throws away already finished work, such as, cache and built artifacts that the most recent run should produce again.
  \item \emph{Latest result vs History} --- The last result is the most important since is the one that will end up in the next stage of the SDLC. However, a branch whose intermediate runs never end up cannot be reviewable, bisected and auditable.
  \item \emph{Abrupt Interruption vs Graceful Interruption} --- A step that take actions against target, for example mid-way through the deployment, leave the pipeline in state where interrupting the run is more damaging than letting it finish.
\end{itemize}

\field{Solution} There is a need to have concurrency groups derived from the branches or pull request developers are working on where a pipeline run should be assign to. Design then your pipeline so that, when a new pipeline run needs to start, previous runs based on now-outdated input are stopped. Irreversible side effects jobs and steps should be excluded from cancellation~\cite{khatami2024catching,bouzenia2024resource}.

\field{Consequences}
\begin{itemize}\setlength\itemsep{1pt}\setlength\topsep{2pt}
  \item[\pro] A developer do not need to wait for the other runs to be executed to get its feedback.
  \item[\pro] A developer receives a single result rather than multiple results from the different commits.
  \item[\con] Only the last run is judged, leaving the intermediate pipeline runs without verdict, which can hide developer feedback.
\end{itemize}

\field{Related patterns} \emph{Single-Responsibility Steps} determine where a pipeline run can be gracefully interrupted since a step compasses the evaluation to determine the risk for a step to be canceled.

\pattern{wct}{Workflow Component Template}

\field{Context} Multiple similar jobs with only partially varying implementations.

\field{Problem} A maintainer create a job/step to a set of actions for a service, then copies it for another service, and so on. Later something need to change across all service in that job/step, meaning that every job/step created for each service need to be updated. Fully defining each implementation of similar jobs will result in unnecessary repeated code. This is because: Each job shares identical logic. This logic has to be defined for each full implementation of the job.

\field{Forces}
\begin{itemize}\setlength\itemsep{1pt}\setlength\topsep{2pt}
  \item \emph{One place to edit vs Readability} --- Maintaining a single definition is better when logic is shared shared. Nevertheless, reading the job can no longer show what it have run decreasing readability of that pipeline definition and execution. 
  \item \emph{Variants vs Readability} --- Team want a template to cover every job that is nearly the same, but each job variant add a condition or parameter turning the template into a control flow harder to read.
  \item \emph{Big Bang vs Blast Radius} --- We want to propagate the change at once but a mistake reach the consumers as fast as the propagation.
\end{itemize}

\field{Solution} Define a template containing all the logic each implementation will share. Implement the template in a way that allows its behavior to be adjusted by specific implementations. Define each specific implementation so that it exclusively adjusts the behavior of relevant sections of its template~\cite{duvall2011cdpatterns,duvall2010cipatterns}.

\field{Consequences}
\begin{itemize}\setlength\itemsep{1pt}\setlength\topsep{2pt}
  \item[\pro] Shared logic is fixed at once rather than for each implementation.
  \item[\pro] Implementation that should behave alike can share code.
  \item[\pro] Implement a similar service is faster.
  \item[\con] Fixing a step is easy but it can add defects to other implementations.
\end{itemize}

\field{Related patterns} \emph{Single-Reponsibility Steps} are the pieces that can be used within templates.

\pattern{ebt}{Event-based Triggering}

\field{Context} A software project has different CI/CD workflows corresponding to different action that need to be executed depending on certain events --- Pull Requests, commits, issue activities, tag, and many others --- These events happen at unpredictable times nor their frequency is known in advance.

\field{Problem} There's no suitable fixed schedule for unplanned events. Scheduled workflow runs either wastes resources or leaves events unprocessed. Relying on developers to manually trigger workflows means execution depends on memory and availability. How can the appropriate workflow be started promptly and reliably every time an event occurs?~\cite{yasa2023developers,vassallo2019automated,duvall2010cipatterns}

\field{Forces}
\begin{itemize}\setlength\itemsep{1pt}\setlength\topsep{2pt}
  \item \emph{Feedback Speed vs Resource Consumption} --- The sooner the developer has feedback the better for developer act promptly. However, every run uses computational resources and pipeline can be triggered faster than they complete.
  \item \emph{Automation vs Control} --- Events that trigger specific workflows guarantee that no manual execution is required, but it also remove the ability to choose when a particular change don't need to trigger a workflow.
  \item \emph{Trigger granularity vs Configuration Simplicity} --- The granularity of the events that trigger a workflow avoid unnecessary executions, however each aditional event-based condition added to the workflow configuration file make it hard to understand.
\end{itemize}

\field{Solution} Adjust your CI/CD so that certain workflows are subscribed to specific events. Whenever a matching event to a workflow occurs, that workflow is triggered~\cite{duvall2010cipatterns,duvall2007continuous}.

\field{Consequences}
\begin{itemize}\setlength\itemsep{1pt}\setlength\topsep{2pt}
  \item[\pro] Feedback arrive fast for the developer to act promptly.
  \item[\pro] Trigger is no longer human dependent for the day-to-day activities.
  \item[\con] Workflow runs become unpredictable, therefore infrastructure need to provide capacity for peaks that might occur.
  \item[\con] The number of executions that an event trigger cannot be controlled, therefore it can generate computational waste, unless interruption mechanisms are implemented.
\end{itemize}

\field{Related patterns} \emph{Interrupt Outdated Pipeline Runs} is complementary to this pattern in order to solve one of its disadvantages, the number of executions that an event trigger. Therefore, any workflow that is trigger based on the same event can be suppressed. Also, \emph{Scheduled-Triggering} has can cover specific workflow runs that this pattern cannot serve and \emph{Manual Triggering} where it can keep the human able to trigger when automatic is not desirable.

\pattern{mt}{Manual Triggering}

\field{Context} A software system implementing workflows that need only be triggered sporadically and ad hoc.

\field{Problem} Some workflows need only be run at irregular intervals. Sometimes the triggering conditions of these workflows cannot be defined programmatically. How should such workflows be triggered?~\cite{yasa2023developers,vassallo2020configuration,duvall2007continuous,duvall2011cdpatterns,bajpai2024cicd}

\field{Forces}
\begin{itemize}\setlength\itemsep{1pt}\setlength\topsep{2pt}
  \item \emph{Flexibility vs Reproducibility} --- Human control of the workflow parameters and execution time make the workflow serve different proposes and fit a lot of situations, however make every run potentially different. 
\end{itemize}

\field{Solution} Adjust your CI/CD infrastructure so that certain workflows are triggered manually.

\field{Consequences}
\begin{itemize}\setlength\itemsep{1pt}\setlength\topsep{2pt}
  \item[\pro] Gives a deliberate control over a workflow.
  \item[\pro] Resource consumption is not unpredictable.
  \item[\con] Workflow execution depend on a human.
\end{itemize}

\field{Related patterns} This pattern cover the cases \emph{Event-Based Triggering} and \emph{Scheduled Triggering} cannot cover. It can act as the vehicle to manual rollback which implements the \emph{Rollback Strategy}.

\pattern{st}{Scheduled Triggering}

\field{Context} A software system implementing workflows that need to be triggered at regular intervals.

\field{Problem} Some workflows must be executed at regular intervals (for example, workflows related to pipeline monitoring or to trigger test automation). How can we best ensure such workflows are triggered regularly?

\field{Forces}
\begin{itemize}\setlength\itemsep{1pt}\setlength\topsep{2pt}
  \item \emph{Detection Latency vs Resource Consumption} --- Shorter interval runs may reveal problems fast, however it increase resource consumption.
\end{itemize}

\field{Solution} Adjust your CI/CD infrastructure so that certain workflows are triggered automatically at predefined intervals~\cite{duvall2010cipatterns,bajpai2024cicd,bouzenia2024resource}.

\field{Consequences}
\begin{itemize}\setlength\itemsep{1pt}\setlength\topsep{2pt}
  \item[\pro] .
  \item[\con] An error introduced after a scheduled run stays undercover.
  \item[\con] .
\end{itemize}

\field{Related patterns} This pattern is the alternative to \emph{Event-Based Triggering}, has can be used to execute work that is too slow or costly to run on every change. \emph{Manual Triggering} is the other pattern to be used for human decision workflow runs.

\pattern{filter-external-input}{Filter External Input}

\field{Context} The workflow uses input generated by an entity that it does not have complete control over.

\field{Problem} Workflows can be compromised by third-parties. This is because: If any part of a workflow is compromised, the entire workflow may be compromised as well. A software system cannot guarantee the integrity of what falls beyond its control. External input can potentially be compromised (intentionally or unintentionally) by third parties~\cite{li2024toward,huang2025revisiting}.

\field{Forces}
\begin{itemize}\setlength\itemsep{1pt}\setlength\topsep{2pt}
  \item \emph{Security vs Sanitization} --- By filtering the external input we can improve the security of a workflow, however, excessive filtering may hide useful information or change the workflow behaviour.
  \item \emph{}
\end{itemize}

\field{Solution} Validate, filter and sanitize external input, before process the workflow in order to ensure no potentially harmful data is used by the workflow.

\field{Consequences}
\begin{itemize}\setlength\itemsep{1pt}\setlength\topsep{2pt}
  \item[\pro] Reduce the risk of external input compromises the workflow.
  \item[\pro] Improve the overall reliability of the workflow.
  \item[\con] Excess filtering may reject useful inputs.
  \item[\con] The overall complexity of the workflow increases has it introduces additional processing.
\end{itemize}

\field{Related patterns} \emph{Deterministic Dependencies} complements this pattern as it implements a way to filter data that enters a workflow by pinning the version that can be downloaded. \emph{Permission Management} can act as a limiter to the damage an external injection can do.

\pattern{notifications-mechanism}{Notification Mechanisms}

\field{Context} A software system development or maintained by a team or a group of teams.

\field{Problem} Issues detected in a pipeline may go unnoticed and unfixed for prolonged periods of time. This is because: During the execution of a pipeline, there are often issues which require manual input to be resolved (for example: fixing code that fails builds or tests). Fixing these issues quickly is essential to maximizing the efficiency of a pipeline. These issues often reveal themselves at unpredictable times. Developers are not constantly monitoring detected issues or the pipeline behaviour~\cite{vassallo2019automated,duvall2007continuous,duvall2011cdpatterns,duvall2010cipatterns,bajpai2024cicd}.

\field{Forces}
\begin{itemize}\setlength\itemsep{1pt}\setlength\topsep{2pt}
  \item \emph{Actionability vs Disruption} --- Timely notifications improve the responsiveness required for a developer to act promptly, nevertheless, numerous alerts that do not reveal actionable steps can be distracting and contribute to fatigue.
\end{itemize}

\field{Solution} Implement notification mechanisms to alert relevant parties about a workflow behavior that require human to look into its result~\cite{duvall2011cdpatterns,duvall2010cipatterns,bajpai2024cicd}.

\field{Consequences}
\begin{itemize}\setlength\itemsep{1pt}\setlength\topsep{2pt}
  \item[\pro] Reduces the issue resolution time.
  \item[\pro] Has relevant parties receive feedback about the pipeline behavior, it increases the pipeline efficiency.
  \item[\pro] It can enhance visibility of pipeline failures and the actions that are required.
  \item[\con] The phenomenon of alert fatigue can be observed, resulting from excessive or poorly targeted notifications.
  \item[\con] This mechanisms depends on the integration with the target app that receive notification, this need further configuration and may add costs.
\end{itemize}

\field{Related patterns} The notifications that this pattern delivers can be produced by \emph{Self-monitoring Pipelines}. This pattern also enables \emph{Fail Fast} as a failing fast is only useful if someone gets notified. Furthermore, it's important for the \emph{Rollback Strategy} has rollbacks need to be notified.

\pattern{rollback-strategy}{Rollback Strategy}

\field{Context} A software system that is delivered to stakeholders through large, infrequent releases.

\field{Problem} A previous version of a software system may be superior to its most recent version. This is because: A software system most often requires numerous changes in order to adapt to the ever-shifting needs of its users. These changes should be applied as quickly as possible in order for users to have access to up-to-date functionalities. There always exists a possibility that a change will introduce unexpected issues into a software system that make its current version worse than the previous one~\cite{duvall2007continuous,duvall2011cdpatterns}.

\field{Forces}
\begin{itemize}\setlength\itemsep{1pt}\setlength\topsep{2pt}
  \item \emph{Time to recover vs Complexity} --- Implementing fast rollback mechanisms have the effect of reducing downtime and disruption. However, it should be noted that such mechanisms may require additional infrastructure, processes and version control.
  \item \emph{Frequent releases vs stability} --- Frequent releases deliver faster value customers,  but each new delivery may hurt software stability increasing the the risk of introducing bugs.
\end{itemize}

\field{Solution} Structure the software system in a way where it can easily be restored to an earlier version and subsequently distributed to its users~\cite{duvall2007continuous,duvall2011cdpatterns}.

\field{Consequences}
\begin{itemize}\setlength\itemsep{1pt}\setlength\topsep{2pt}
  \item[\pro] Reduces the risk of introducing bugs to customers.
  \item[\pro] Reduces deployment failures.
  \item[\pro] Increase release confidence.
  \item[\con] Requires versioning and rollback processes.
\end{itemize}

\field{Related patterns} \emph{Notification mechanism} has its important to alert that software has been rollback. Recovering from a fault can be done in different ways, with one such option being the use of a \emph{Fast-lane for Hot-fixes}.

\pattern{task-parallelization}{Task Parallelization}

\field{Context} Pipeline with independent steps.

\field{Problem} Executing a pipeline as a single continuous operation tends to lead to an under- performance of the hardware used to run the pipeline. This is because most modern hardware has been designed to perform best when executing multiple operations at once. 

\field{Forces}
\begin{itemize}\setlength\itemsep{1pt}\setlength\topsep{2pt}
  \item \emph{Speed vs Complexity} --- Executing tasks in parallel increase pipeline speed, however it increase the complexity to design and maintain pipeline steps coordination. 
  \item \emph{Resource Utilization vs Resource Competition} --- Executing tasks in parallel increase the resource utilization, but it increase the change of race to conditions where pipeline compete for resources.
\end{itemize}

\field{Solution} Design your pipeline to harness the potential of the hardware it relies on by parallelizing its execution~\cite{urdih2025architectural,bajpai2024cicd}. \footnote{Many CI/CD providers partially abstract away the implementation of this pattern by automatically using parallelization whenever possible}

\field{Consequences}
\begin{itemize}\setlength\itemsep{1pt}\setlength\topsep{2pt}
  \item[\pro] Reduce the execution time.
  \item[\pro] Enhance the usage of the available resources.
  \item[\con] Increased complexity for the pipeline steps dependencies.
  \item[\con] Parallelization design must be well design to have race to condition.
\end{itemize}

\field{Related patterns} The pattern relies on the division of workflow into smaller (\emph{Single-Responsibility Steps}), independent units that can be executed simultaneously.

\pattern{prioritize-failed-jobs}{Prioritize Failed Jobs}

\field{Context} Pipeline with independent jobs.

\field{Problem} Often, resources are wasted by re-running a pipeline only for it to fail on the same job as its previous execution. This is because: Running a pipeline consumes resources. Normally, a job failing leads to the entire pipeline execution failing. When a job causes a pipeline to fail, the next pipeline execution with similar input is more likely to fail on that job than on other jobs. The predefined order of a job in a pipeline is not directly related to how likely it is to fail for a given input. How can a pipeline minimize the resource spent on failing subsequent pipeline?

\field{Forces}
\begin{itemize}\setlength\itemsep{1pt}\setlength\topsep{2pt}
  \item \emph{Optimal Failure Ordering vs Dependencies Constraints} --- The jobs/steps that are more likely to fail should run sooner, however the workflow steps dependencies limit how jobs/steps can be ordered. 
\end{itemize}

\field{Solution} Implement a mechanism that orders job execution so that jobs that failed during previous executions of a pipeline are run earlier in the pipeline whenever possible~\cite{bouzenia2024resource}. \footnote{Prioritize Failed Jobs is dependent on the existence of an implementation of Fail Fast, where the pipeline stops when a job fails.}

\field{Consequences}
\begin{itemize}\setlength\itemsep{1pt}\setlength\topsep{2pt}
  \item[\pro] Reduce resource consumption.
  \item[\pro] Shorten feedback loops.
  \item[\con] Require pipeline history.
  \item[\con] Previous failures may not predict new ones, limiting this pattern efficiency.
\end{itemize}

\field{Related patterns} The idea behind \emph{Fail Fast} is to stop early on when things go wrong, and that if something will probably fail anyway, then you should make that clear right at the start.

\pattern{fast-lane-for-hotfixes}{Fast-lane for Hot-fixes}

\field{Context} A software system that is delivered to stakeholders through small and frequent releases.

\field{Problem} Forcing fixes for serious issues to pass through the same deployment pipeline as regular changes can lead to greater damages due to a longer time until deployment. This is because: No matter how well designed and extensive a deployment pipeline is, it is always possible for serious issues to slip through undetected. A serious issue in a software system needs to be resolved as soon as possible in order to minimize the extent of the damage caused by it. CI/CD pipelines associated with regular changes to the software system are normally quite extensive and consequently time consuming~\cite{duvall2011cdpatterns,duvall2010cipatterns}. 

\field{Forces}
\begin{itemize}\setlength\itemsep{1pt}\setlength\topsep{2pt}
  \item \emph{Speed vs Security} --- Implement a fast lane enable developers to solve issues promptly; Nevertheless, it can introduce security breaches as the full CI/CD pipeline is skipped.
  \item \emph{Usage vs Pipeline Design} --- Fast lane are optimal and ideal for developer to act rapidly; However, this mechanism is tightly coupled to a pipeline design and process which need configuration and maintenance.
\end{itemize}

\field{Solution} Implement a separate deployment pipeline to be used by hot-fixes. Use this pipeline very sparingly and exclusively for urgent adjustments.

\field{Consequences}
\begin{itemize}\setlength\itemsep{1pt}\setlength\topsep{2pt}
  \item[\pro] Solve production issues fast.
  \item[\pro] Flexibility of the workflow.
  \item[\con] Full CI/CD pipeline is skipped potentially introducing security problems as steps are skipped.
  \item[\con] Maintenance is higher.
\end{itemize}

\field{Related patterns} \emph{Rollback Strategy} is an alternative to this pattern and it reverts instead of fixing froward.

\pattern{build-different-system-components-separately}{Build Different System Components Separately}

\field{Context}  A software system which has been (or which can be) divided into multiple loosely coupled modules.

\field{Problem}  By always rebuilding the entire software system independent of the changes made to it, computational resources will be wasted. This is because: Building a software system consumes computational resources relative to its size. Some changes only affect certain parts of a software system. A part of a software system is always smaller than the whole software system.

\field{Forces}
\begin{itemize}\setlength\itemsep{1pt}\setlength\topsep{2pt}
  \item \emph{Resource Optimization vs Configuration} --- Despite the resource optimization that comes from this pattern implementation, has it rebuild only the affected component it increases the configuration effort has it need setup and new services need new configuration. 
\end{itemize}

\field{Solution} Organize your pipeline so that only the parts of your software system affected by a change are rebuilt. Rely on previous build results for unchanged parts~\cite{duvall2007continuous,bajpai2024cicd}.

\field{Consequences}
\begin{itemize}\setlength\itemsep{1pt}\setlength\topsep{2pt}
  \item[\pro] Resource optimization.
  \item[\con] Configuration become harder.
\end{itemize}

\field{Related patterns} \emph{Task parallelization} enable even faster build if services can be built independently. 

\pattern{self-monitoring-pipelines}{Self-monitoring Pipelines}

\field{Context} CI/CD pipelines whose health is critical for the correct functioning of a software system.

\field{Problem} It is often difficult and inefficient to monitor CI/CD pipelines externally. This is because: To accurately identify characteristics that only reveal themselves during pipeline execution, a lot of information about pipeline runs is often required. Retrieving certain information from a pipeline’s runs from outside that pipeline is difficult, sometimes even impossible.

\field{Forces}
\begin{itemize}\setlength\itemsep{1pt}\setlength\topsep{2pt}
  \item \emph{Observability vs Simplicity} --- The data of interest is only observable from within the pipeline; however, the placement of monitoring logic there results in the CI/CD being coupled with its own instrumentation.
\end{itemize}

\field{Solution} Leverage CI/CD values by implementing pipeline metric collection within your pipeline, and implement methods (manual or automatic) to analyze these collected metrics for issues~\cite{duvall2007continuous,duvall2011cdpatterns,duvall2010cipatterns}.

\field{Consequences}
\begin{itemize}\setlength\itemsep{1pt}\setlength\topsep{2pt}
  \item[\pro] Uniform methods on every pipeline run making trend analysis and regression detection easier.
  \item[\pro] Per-step observable data.
  \item[\pro] No external infrastructure, metrics are produced and collected directly in the pipeline.
  \item[\con] Every pipeline execution time increase.
  \item[\con] Maintenance increase.
  \item[\con] Shared abstraction must be implemented, so no duplicated code is generated across different pipelines.  
\end{itemize}

\field{Related patterns} \emph{Single-Responsibility Steps} and \emph{Workflow Component Template} ensure that the observability mechanisms are implemented without being duplicated across pipelines. \emph{Notification Mechanisms} consumes the output generated of self-monitored pipelines.

\pattern{permission-management}{Permission Management}

\field{Context} A software system whose contributors have varying levels of clearance.

\field{Problem} Entities can potentially use the CI/CD pipeline to access and tamper with assets they are not cleared to access. This is because: Due to the overarching nature of CI/CD pipelines, which reaches across most facets of a software system, a pipeline will normally have access to, and sometimes be able to modify, a software system's sensitive information. CI/CD pipelines are generally used by every contributor of a software system. Some contributors of a software system may not have the proper clearance to access certain assets of said system~\cite{khatami2024catching,huang2025revisiting}.

\field{Forces}
\begin{itemize}\setlength\itemsep{1pt}\setlength\topsep{2pt}
  \item \emph{Autonomy vs Control} --- Unrestricted pipeline management enable developers to act promptly and adapt their CI/CD to the circumstances that their product might need to adapt to. However the lack of control can lead to information exposure or malicious tampering.
  \item \emph{Granularity vs Maintenance} --- Fine-grain permissions enable access only to the resources developers need, yet comes with the design and maintenance cost aligned with the organization.
\end{itemize}

\field{Solution} Implement and enforce a CI/CD pipeline permission management system adjusted to the structure of the organization that owns the pipelines~\cite{duvall2011cdpatterns,bajpai2024cicd,khatami2024catching,huang2025revisiting}.

\field{Consequences}
\begin{itemize}\setlength\itemsep{1pt}\setlength\topsep{2pt}
  \item[\pro] Developers can only reach the resources that they have permission to.
  \item[\pro] Permissions can mirror organization structure, enabling audit and traceability.
  \item[\pro] The blast radius of a potential harmful and compromising account is limited to the pipeline permissions.
  \item[\con] Permission matrix model must be maintained.
  \item[\con] Overly restricted pipelines can hamper software delivery.
  \item[\con] The permission system become a single point of failure.
\end{itemize}

\field{Related patterns} This pattern complements the \emph{four eyes principle} has it define how many people must agree and this pattern on who act.

\pattern{four-eyes}{The Four Eyes Principle}

\field{Context} A software system with multiple contributors.

\field{Problem} Some important quality aspects cannot be automatically verified by a pipeline. This is because: Many aspects of software quality cannot be automatically evaluated, either because there exists no known method of automatic evaluation for them or because said method is prohibitively expensive. These aspects of software quality are oftentimes still relevant and should therefore be kept above certain thresholds.

\field{Forces}
\begin{itemize}\setlength\itemsep{1pt}\setlength\topsep{2pt}
  \item \emph{Coverage vs Human Judgment} --- Automation is repeatable and cheap, compared to human review, but only a human can assess certain design, and architectural intents. 
  \item \emph{Autonomous vs Manual Gate} --- CI/CD is design to be autonomous and run unattended, yet a manual gate deliberately includes a step that requires a human intervention and suspends the pipeline execution until an action is taken.
\end{itemize}

\field{Solution} Require another human collaborator to manually inspect/review the quality aspects of submitted changes which the pipeline cannot before admitting said changes into the repository. The approval method come from the CI/CD mechanisms so the verdict is a pipeline signal like any other automation in said pipeline~\cite{duvall2011cdpatterns}.

\field{Consequences}
\begin{itemize}\setlength\itemsep{1pt}\setlength\topsep{2pt}
  \item[\pro] Human intervention is based on automation that has already been implemented for the pipeline. Consequently the human decision is augmented by this automation.
  \item[\pro] Pipeline become source of auditing processes with the information who admitted which change.
  \item[\pro] It is not possible for a single contributor to make effective changes alone.
  \item[\con] The manual gate hamper flow, as it introduces a wait state only a human can approve.
  \item[\con] Time-to-market pressure can lead the approval gate into rubber-stamping making the gate redundant.
  \item[\con] It is imperative that the gate is protected in order to ensure that only developers who have been granted the appropriate authorization are able to remove it or bypass it.
\end{itemize}

\field{Related patterns} \emph{Permission Management} defines who is allowed to act, complementing this pattern. It should be noted that this pattern approach is in tension with the \emph{Fast-lane for Hot-fixes}, which may require the relax of the review period to facilitate the deployment of urgent fixes.

\pattern{deterministic-dependencies}{Deterministic Dependencies}

\field{Context} A software system that uses external dependencies.

\field{Problem} The behavior of software systems using external dependencies can be altered by changes outside of the software system's control. This is because: The evolution of external dependencies is, by definition, completely or at least mostly outside of the control of those responsible for the pipelines that use them. Maintainers of external dependencies can modify interfaces or functionality, introduce errors, or otherwise alter an external dependency. If a dependency changes, so do the software systems that depend on it~\cite{vassallo2020configuration,khatami2024catching,li2024toward,pan2024ambush,huang2025revisiting}.

\field{Forces}
\begin{itemize}\setlength\itemsep{1pt}\setlength\topsep{2pt}
  \item \emph{Reproducibility vs Staleness} --- Pinning dependencies to a specific version ensures that every pipeline run uses the same dependency version, however this can result in the system becoming stale if no policy is in place to manage the updating of dependencies.
  \item \emph{Updates vs Control} --- By using tag or references that floats, pipeline dependencies can automatically improve or get fixes, yet developers loose the control over what actually their pipeline do.
  \item \emph{Maintenance vs Debugging} --- Maintain dependencies references up to date can be expensive, but chasing failures caused by upstream changes can be more expensive.
\end{itemize}

\field{Solution} When referencing an external dependency, make sure the reference used points to a static instance, thereby ensuring that your pipeline is independent of the dependency's evolution~\cite{khatami2024catching,li2024toward,huang2025revisiting}.

\field{Consequences}
\begin{itemize}\setlength\itemsep{1pt}\setlength\topsep{2pt}
  \item[\pro] CI/CD pipelines become reproducible.
  \item[\pro] System changes are not assigned to third parties.
  \item[\con] Updates to the pipeline dependencies are no longer automatic.
  \item[\con] Accumulating updates makes the upgrade riskier.  
  \item[\con] You must keep a mirror of a version of the dependency, so if the reference disappear you still have the possibility to use that version.
\end{itemize}

\field{Related patterns} The \emph{Filter External Input} pattern can complement this pattern as it addresses two key areas: firstly, the presence of untrusted inputs, and with this pattern the presence of untrusted code.

\pattern{selective-testing}{Selective Testing}

\field{Context} Pipeline with tests targeting only subsections of a software system.

\field{Problem} Running all tests independent of the changes made to the software system will likely waste resources. This is because: Running tests inside a pipeline consumes resources. Committed code should be tested. Code changes need not affect the whole software system, many sections of code may remain unchanged. If a section of code does not change, the results of tests verifying said section also will not change.

\field{Forces}
\begin{itemize}\setlength\itemsep{1pt}\setlength\topsep{2pt}
  \item \emph{Coverage vs Resource Usage} --- The implementation of a comprehensive testing strategy is an essential component of ensuring system coverage. However, it should be noted that this approach entails an investment of time and resources, despite the possibility that the outcomes may remain unchanged.
  \item \emph{Feedback Speed vs Confidence} --- A concise, focused run ensures the flow of developers, while a comprehensive run provides the most robust coverage of the system's integrity.
  \item \emph{Selection accuracy vs cost} --- The selection of the tests that will run needs an analysis, which is a process that must be undertaken, managed and trusted.
\end{itemize}

\field{Solution} Adjust your pipeline so that, when code changes are processed, only the tests whose result is affected by these changes are run~\cite{urdih2025architectural}.

\field{Consequences}
\begin{itemize}\setlength\itemsep{1pt}\setlength\topsep{2pt}
  \item[\pro] CI/CD consume less resources.
  \item[\pro] Failures diagnosis are easier to detect.
  \item[\pro] CI/CD can run often.
  \item[\con] Incorrect test selection can result in the skipping of essential testing, thereby facilitating regressions.
  \item[\con] The test selection need extra pipeline configuration, which need to be maintained.
  \item[\con] Different runs execute different tests, meaning a green pipeline no longer means that it has executed the same thing.
\end{itemize}

\field{Related patterns} \emph{Single-Responsibility Steps} enable this pattern to use the granularity that determines which tests runs. \emph{Build Different System Components Separately} highlights the change, therefore it facilitates the selection of which tests should be executed.

\bibliographystyle{splncs04}
\bibliography{references}

\end{document}